\documentstyle[12pt]{article}
\newcommand{\be}{\begin{equation}}
\newcommand{\ee}{\end{equation}}
\newcommand{\ba}{\begin{eqnarray}}
\newcommand{\ea}{\end{eqnarray}}
\newcommand{\bb}{\begin{thebibliography}}
\newcommand{\eb}{\end{thebibliography}}

\begin{document}

\begin{center}
\vspace*{1.cm}
{ \Large \bf DIPOLE AND QUADRUPOLE MOMENTS
  OF MIRROR NUCLEI $^{8}$B AND $^8$Li } \\
\vspace*{0.2cm}

{ \large\bf   G.Kim \\ }
{ \it Institute of Nuclear Physics, Tashkent, Uzbekistan }  \\
{ \large\bf   R. R. Khaydarov \\ }
{ \it Institute of Nuclear Physics, Tashkent, Uzbekistan }  \\
{ \large\bf Il-Tong Cheon } \\
{ \it Department of Physics, Yonsei University, Seoul 120 -- 749, Korea } \\
{ \large\bf F.A. Gareev } \\
{ \it JINR, Dubna, Russia } \\

\end{center}

Magnetic dipole and electric quadrupole moments of the mirror nuclei $^8Li$
and $^8B$ are analysed in the framework of the multiparticle shell model by
using two approaches :
i) the one-particle spectroscopic factors and
ii) the one-particle fractional parentage coefficients.
These two approaches are compared both each to other and with
a microscopic multicluster model.
The one-particle nucleon states are calculated taking  into
account the continuum by the method of the expansion of the
Sturm - Liouville functions.
The experimental magnetic and quadrupole moments of  $^8Li$ and $^8B$
are reproduced well by using fractional  parentage  coefficients
technique.
The root mean-square radii and the radial  density  distributions
are obtained for these nuclei.\\

{\large \bf 1.  Introduction}\\

{\large

 From a known proton-rich
nuclei the nucleus $^8$B has an extremely low
 binding energy of the valence proton and is one of the best candidates for
 the proton-halo nucleus. The $^8$B proton-halo problem was considered in many
 experimental and theoretical works (see [1] and the references listed there).
 At present wide experimental material has been accumulated: the total
 interaction cross section of $^8$B with the different nuclei in the wide
 range of the incident energies, the quasielastic scattering data on  different
 targets, the fragmentation data of $^8$B to $^7$Be and proton, as well as
 the measurement of the electric quadrupole moment of $^8$B [2]. And although
 the latest publications [1] can explain the existing experimental data by
 a proton halo in $^8$B, it is necessary to understand, whether spatially
 extended density distribution of the loosely bound proton is the property of
 the structure $^8$B itself or the introduction of the halo is consequence of
 our insufficient knowledge of the reaction mechanisms.

 Of course, when the investigations of nucleus-nucleus collisions are performed,
 the questions of structure of the interacting nuclei and the reaction dynamics
 are closely interlaced. Therefore, it would be the ideal variant to have
 the form factors of elastic electron scattering on $^8$Li and $^8$B.
 As the electron-nucleus interaction is, mainly, electromagnetic and,
 in principle, well known one, the most precise information on the nuclear
 structure may be derived from the analysis of these form factors together with
 the data of electric and magnetic moments for these nuclei. And if it is
 realized, then the trustworthy of the reaction mechanisms for different
 processes can be deduce. As far as the obtaining of electron scattering data
 on unstable nuclei - is the problem of the future investigations [3], one has
 to carry out the careful analysis of the static data: electric quadrupole and
 magnetic dipole moments of nuclei $^8$Li and $^8$B. The description of these
 data depends completely on the structure of these nuclei.

 A "nucleon halo" - is a new form of nuclear matter, characterized
 by low and nonuniform density distribution. The description of the halo nuclei
 needs the extra caution. It is necessary to pay the special attention to
 the following, very  important, factors :
 I) the asymptotic behavior of the wave function of the valence nucleon;
 II) the correct consideration of the continuum effects;
 III) taking into account of the nucleon associations or the clustering
 phenomena;
 IV) the  exact treatment of the Pauli principle or
 the antisymmetrization effects.
 At once one should point out that all these factors are intercorrelated.
 The different models of the resonating-group method (RGM)
 (or the generator-coordinate method (GCM)) [4,5] are the most promising ones
 from the point of view of the clustering phenomena and the Pauli principle.
 It may be the nucleon association model [6] or the microscopic multicluster
 model intensively developing now using the stochastic variational method [7].
 These models assume the wide use of powerful computer facilities. Moreover,
 so far as in these models the basis function are chosen in Gaussian form,
 for the achievement of the correct asymptotic form of the wave functions and
 for the taking into account the continuum (the high partial wave channels)
 it is necessary to take a very large basis of the trial functions.
 The calculation becomes too complicated and computer time consuming.
 In spite of all respect to the valuable results, obtained by such a way,
 one should be point out, that the big heuristic power of the analytical method
 being a characteristic feature for the traditional theoretical physics, here
 is reduced noticeably.

 In present work the multiparticle shell model [8,9].
 for the description of magnetic dipole and electric quadrupole moments
 $^8$Li and $^8$B. The Pauli principle is treated exactly in this model.
 For obtaining the correct asymptotic behavior in our approach we used
 the single-particle wave functions calculated in the Woods-Saxon potential
 both with and without the spin-orbit interaction [10]. If the energy and
 the wave function of the quasi-bound $1p_{1/2}$ - state are calculated
 taking into account the spin-orbit interaction, then the method of
 the expansion of the Sturm-Liouville functions is used, i.e the calculations
 are performed with an appropriate treatment of continuum effects [11,12].
 Concerning to the clusterization, in the shell model the construction
 of the shell wave function by itself is connected with the associative
 structure of the nuclear state. For instance, in the Young's scheme
 [f1f2f3]=[444], corresponding to the structure from three $\alpha$-clusters,
 the attraction within a separate line (i.e the attraction of nucleons inside
 the association) and the repulsion between nucleons of the different clusters
 (lines) dominates. Let us mention also the well-known fact, that if
 the harmonic oscillator size parameters of the wave functions of the internal
 and relative motions are equal, then there is the identity between the wave
 functions of RGM and the shell model. If the oscillator parameter of the wave
 function of the relative motion of the clusters differs from that of
 the motion of nucleons inside the cluster, then one can expand the relative
 wave function on the principal quantum number with respect to the intrinsic
 wave functions. Then every term of the expansion will represent the wave
 function of the shell model and the expansion coefficient will define
 the weight of the corresponding configuration into the common wave functions
 of RGM. And this means that such approach allows to determine "non-obviously"
 the influence of continuum on the RGM wave function. This multiparticle shell
 model with the realistic radial wave functions and the correct asymptotic
 behavior, which takes into account exactly of the continuum effects contains
 the aforesaid factors, that is necessary for considering the weakly bound halo
 nuclei.\\

{\large \bf 2.Formalism}\\

 The wave functions for the nuclei with A=8 in ground state are as follows:

\be
{\Psi}_{{\alpha}T}(gr.st)=  \sum_{i}^{}{\beta}_i|(1p)^4[{\lambda}]^{(2T+1)
(2S+1)}L_j>,
\ee
\\
where
${\beta}_i$ -configuration mixing coefficients,   determining  state  weights
with given values of orbital -L,  spin -S and isospin -T moments for Young's
scheme [$\lambda$] in the complete wave function of 4 nucleons in the 1p-shell,
J - the total momentum of nucleus. If the values of coefficients ${\beta}_i$
are known, then it is very convenient to calculate the matrix elements of
the single-particle operators with such wave functions (1), separating
the single-particle states with the fractional parentage coefficients
technique. For the considered nuclei $^{8}B$ and $^{8}Li$ we have --
$J^{\pi}=2^{+},\ T=1$. The nuclear quadrupole and magnetic moments are defined
by the proton-neutron formalism as:

\ba
eQ=\frac{8}{5} \sqrt{\frac{2}{7}} \sum_{{\alpha}=p,n}^{}e_{\alpha}^{eff}
{\lbrack} < \varphi_{1p_{1/2}}^{(\alpha)}(r)|r^2|\varphi_{1p_{3/2}}^{(\alpha)}(r)>
X_{\frac{1}{2}, \frac{3}{2},2}^{({\alpha} )} - \qquad\qquad\qquad\nonumber\\
\nonumber \\
 - <\varphi_{1p_{3/2}}^{(\alpha)}(r)|r^2|\varphi_{1p_{3/2}}^{(\alpha)}(r)>
X_{\frac{3}{2},\frac{3}{2},2}^{({\alpha} )}\rbrack, \qquad\qquad\qquad
\ea

\ba
{\mu}_{Nucl.}= \frac{{2}\sqrt{5}}{15} {\lbrace}{e}{\lbrack}
2X_{\frac{1}{2}, \frac{1}{2},1}^{(p)} +
\sqrt{10}X_{\frac{3}{2}, \frac{3}{2},1}^{(p)} -
2\sqrt{2}X_{\frac{1}{2}, \frac{3}{2},1}^{(p)}
< \varphi_{1p_{1/2}}^{(p)}(r)|\varphi_{1p_{3/2}}^{(p)}(r)>\rbrack + \nonumber \\
\nonumber\\
\sum_{{\alpha}=p,n}^{}\mu_{\alpha}{\lbrack}-X_{\frac{1}{2}, \frac{1}{2},1}^{({\alpha})} +
\sqrt{10}X_{\frac{3}{2}, \frac{3}{2},1}^{({\alpha})} +
4\sqrt{2}X_{\frac{1}{2}, \frac{3}{2},1}^{(\alpha)}
< \varphi_{1p_{1/2}}^{(\alpha)}(r)|\varphi_{1p_{3/2}}^{(\alpha)}(r)>\rbrack \rbrace, \qquad
\ea
\\
 where $\alpha= р$ - for protons and $\alpha=n$ - for neutrons;
\vspace*{0.3cm}

 $e_{\alpha}^{eff}$ - the effective charge of nucleon;
\vspace*{0.3cm}

 ${\mu}_{\alpha}$  - the magnetic moment of nucleon;
\vspace*{0.3cm}

 $\varphi_{1pj}^{(\alpha)}(r)$ - the radial function of nucleon in
 $1p_j$ - state;
\vspace*{0.3cm}

 $X_{j^{'},j,k}^{(\alpha)}$ - the nucleon spectroscopic amplitude, which is
 expressed through \\
\hspace*{2.1cm} the weight coefficients ${\beta}_i$ and the fractional
 parentage coefficients
\hspace*{2.1cm} [9,10];
\vspace*{0.3cm}

 $j и j^{'}$  - the total angular moments of nucleon in the initial and final
 states, respectively;
\vspace*{0.3cm}

 k - transferred value of the total angular moment.
\vspace*{0.3cm}

 The one-particle root-mean-square (rms) radius may be presented in these
 nonations as follows:

\be
{<r_{1l_j}^2>}_{\alpha}^{\frac{1}{2}}= \sqrt {<{\varphi}_{1l_j}^{(\alpha)}(r)
|r^2|{\varphi}_{1l_j}^{(\alpha)}(r)>}.
\ee

 For the nucleon occupation numbers of $1l_j$ - states we have [10]:

\be
{\eta}_{1l_j}^{(\alpha)}=\sqrt{\frac{2j+1}{2}}X_{j,j,0}^{(\alpha)}.
\ee

 Here l=p for 1p-shall, l=s for 1s-shall.

 For the nuclear rms radii ( $R_{r.m.s}^{p}$, $R_{r.m.s}^{n}$ and
 $R_{r.m.s}^{m}$ -- the proton, neutron and matter rms radii, respectively) we
 have following expressions:

 $$ R_{r.m.s}^{p}={ \lbrace { \frac{ {\eta}_{1s_{1/2}}^{p}{<r_{1s_{1/2}}^{2}>}_{p} +
 {\eta}_{1p_{3/2}}^{p}{<r_{1p_{3/2}}^{2}>}_{p}+
 {\eta}_{1p_{1/2}}^{p}{<r_{1p_{1/2}}^{2}>}_{p} } {Z}} \rbrace}^{\frac{1}{2}},
 \ \ \ \ (6a)$$

$$ R_{r.m.s}^{n}={ \lbrace { \frac{ {\eta}_{1s_{1/2}}^{n}{<r_{1s_{1/2}}^{2}>}_{n} +
{\eta}_{1p_{3/2}}^{n}{<r_{1p_{3/2}}^{2}>}_{n}+
{\eta}_{1p_{1/2}}^{n}{<r_{1p_{1/2}}^{2}>}_{n}}{N}}\rbrace}^\frac{1}{2},
 \ \ \ \ (6b) $$

$$ R_{r.m.s}^{m}={ \lbrace { \frac{ Z\cdot(R_{r.m.s}^{p})^2+N\cdot(R_{r.m.s}^{n})^2}
{A} \rbrace}}^{\frac{1}{2}},
 \ \ \ \ \ \ \ \qquad\qquad\qquad\qquad\qquad\quad (6c) $$
\\
 where $Z$ -- the number of protons, $N$ -- the number of neutrons in
 the nucleus with the number of nucleons $A$.\\
 If a nucleus is not strongly excited, then
 $$
 {\eta}_{1s_{1/2}}^{p}=\eta_{1s_{1/2}}^{n}=2.
 $$

 At once it should be noted, if a spin-orbit interaction is not taken into
 account for calculation of the radial wave function in the mean field, then \\

 $\varphi_{1p_{3/2}}^{(\alpha)}(r)= \varphi_{1p_{1/2}}^{(\alpha)}(r)=
\varphi_{1p}^{(\alpha)}(r)$,
\hspace*{0.3cm} $<r_{1p_{3/2}}^{2}>_{\alpha}=<r_{1p_{1/2}}^{2}>_{\alpha}=<r_{1p}^{2}>$ \\

 and the expressions (6а) and (6b) will take the following forms:

\ba
 R_{r.m.s}^{p}={ \lbrace \frac{ 2{<r_{1s}^2>}_p+(Z-2){<r_{1p}^2>}_{p}} {Z}
 \rbrace}^{\frac{1}{2}}, \ \ \ \ \ (7a) \nonumber
\ea

\ba
 R_{r.m.s}^{n}={ \lbrace \frac{ 2{<r_{1s}^2>}_n+(N-2){<r_{1p}^2>}_{n}} {N}
 \rbrace}^{\frac{1}{2}}. \ \ \ \ \ (7b) \nonumber
\ea

 It follows from these expressions and (6с), that if the spin-orbit interaction
 is absent, then the nuclear rms radii are independent from the nuclear
 structure and defined completely by the single-particle rms radii or
 the one-particle radial wave functions.

 The nuclear quadrupole and magnetic moments can be expressed by
 the one-particle spectroscopic factors [13]:

 $$ eQ=5\frac{\sqrt{21}}{21}\sum_{{\alpha}=p,n}^{}e_{\alpha}^{eff}
 \sum_{{J_c}T_c (E_c)}^{}\sum_{j{j^{'}}}{\lbrack
 S_{J_cT_c(E_c),1p_{j^{'}},\alpha}^{JT(0)}\rbrack}^\frac{1}{2}
 {{\lbrack}S_{J_cT_c(E_c),1p_j,\alpha}^{JT(0)}{\rbrack}}^\frac{1}{2}\times $$\\
 $$ <\varphi_{1p_{j^{'}}}^{(\alpha)}(r,E_c)|r^2|\varphi_{1p_j}^{(\alpha)}(r,E_c)>
 {(-1)}^{J_c-\frac{1}{2}}\sqrt{(2j+1)(2{j^{'}}+1)}\ \times $$
 \vspace*{0.3cm}
 \begin{displaymath}
 \hspace*{9.0cm}
 \Biggl \{ \begin{array}{lll}
 1     & \frac{1}{2} & j \\
 j^{'} & 2           & 1  \\
 \end {array} \Biggr \}
 \Biggl \{ \begin{array}{lll}
 J     & j^{'} & J_c \\
 j^{'} & J           & 2  \\
 \end {array} \Biggr \}, \qquad \qquad (8)
 \end{displaymath}
\vspace*{0.5cm}

$${\mu}_{Nucl.}=5\frac{\sqrt{5}}{6} \Biggl\{ e \sum_{J_cT_c(E_c)}^{} {\sum_{jj^{'}}^{}}
{\lbrack S_{J_cT_c(E_c),1p_{j^{'}},p}^{JT(0)}\rbrack}^\frac{1}{2}
{{\lbrack}S_{J_cT_c(E_c),1p_j,p}^{JT(0)}{\rbrack}}^\frac{1}{2}\times $$\\
$$<\varphi_{1p_{j^{'}}}^{(p)}(r,E_c)|\varphi_{1p_j}^{(p)}(r,E_c)>
(-1)^{2j+J_c+\frac{3}{2}}\sqrt{(2j+1)(2{j^{'}}+1)}\ \times $$
 \vspace*{0.3cm}
 \begin{displaymath}
 \hspace*{6.0cm}
\Biggl \{ \begin{array}{lll}
1     & \frac{1}{2} & j \\
j^{'} & 1           & 1  \\
\end {array} \Biggr \}
\Biggl \{ \begin{array}{lll}
J     & j^{'} & J_c \\
j^{'} & J           & 1  \\
\end {array} \Biggr \}\ +\\
\end{displaymath}
\vspace*{0.3cm}

$$\sum_{\alpha=p,n}^{ } {\mu}_{\alpha}\sum_{J_cT_c(E_c)}^{} \sum_{jj^{'}}^{ }
{\lbrack
S_{J_cT_c(E_c),1p_{j^{'}},\alpha}^{JT(0)}\rbrack}^\frac{1}{2}
{{\lbrack}S_{J_cT_c(E_c),1p_j,\alpha}^{JT(0)}{\rbrack}}^\frac{1}{2}\times $$ \\
$$ < \varphi_{1p_{j^{'}}}^{(\alpha)}(r,E_c)|\varphi_{1p_j}^{(\alpha)}(r,E_c)>
{(-1)}^{j+j^{'}+J_c+\frac{3}{2}}\sqrt{(2j+1)(2{j^{'}}+1)}\ \times $$
 \vspace*{0.3cm}
 \begin{displaymath}
 \hspace*{8.5cm}
\Biggl \{ \begin{array}{lll}
1     & \frac{1}{2} & j \\
1     & j^{'}           & \frac{1}{2}   \\
\end {array} \Biggr \}
\Biggl \{ \begin{array}{lll}
J     & j & J_c \\
j^{'} & J           & 1  \\
\end {array} \Biggr \} \Biggr\}, \qquad\qquad\qquad (9)
\end{displaymath}

 where $S_{J_cT_c(E_c),1p_j,\alpha}^{JT(0)}$ is the spectroscopic factor of
 the separation of the nucleon (proton - $\alpha$= р or neutron - $\alpha$ =n)
 from 1p - shell with the total angular moment $j$ from nucleus $A$ in
 the ground state with the quantum numbers $JТ$, when the residual nucleus
 (A-р or A-n) is in the state with energy $E_c$ and the quantum numbers
 $J_cT_c$.

 The expression (8) is the detailed presentation of expression (2) from
 the work [13], where the shell-model of Cohen-Kurath and Millener-Kurath
 (C-K and M-K) for the p-shell nuclei with the effective interaction CKPOT was
 used. The radial single-particle wave functions were calculated in
 the Woods-Saxon (WS) potential taking into account the centrifugal and Coulomb
 (for proton) potentials and neglecting the spin-orbit interaction. All the core
 exited states of A-1 nucleus were taken into account. For each core exited
 state with the excitation energy $E_c$, the separation energy of the valence
 nucleon $S_{\nu}$ was determined by the formula

\setcounter{equation}{9}
\be
 S_{\nu}=B(A)-B(A+1)+E_c \ ,
\ee

 where B(A) is the binding energy of the nucleus A.

 The valence single-particle wave function ${\varphi}_{1pj}^{(\alpha)}(r,E_c)$
 was calculated by adjusting the depth of the WS potential to obtain
 the separation energy for each core exited state.

 The expression (4) for the one-particle rms radius is rewritten as:

\be
{<r_{1pj}^2>}_{\alpha,E_c}^{\frac{1}{2}}=
\sqrt {<{\varphi}_{1pj}^{(\alpha)}(r,E_c)
|r^2|{\varphi}_{1pj}^{(\alpha)}(r,E_c)} \ .
\ee
\vspace*{0.3cm}

 And for the nuclear rms radii we obtain:

$$ R_{r.m.s}^p =  \lbrace \frac{1}{Z} \lbrack
 2{<r_{1s_{1/2}}^2>}_p+ 5\frac{\sqrt{5}}{24}
 \sum_{J_cT_c(E_c)}^{}S_{J_cT_c(E_c),1p_{3/2},p}^{JT(0)}
 {<r_{1p_{3/2}}^2>}_{p,E_c} + $$ \\
$$\hspace*{6.5cm}
 5\frac{\sqrt{10}}{24} {\sum_{J_cT_c(E_c)}^{}}
 S_{J_cT_c(E_c),1p_{1/2},p}^{JT(0)} {<r_{1p_{1/2}}^2>}_{p,E_c} \rbrack
 \rbrace ^{1/2}\ , \qquad (12a) $$
\vspace*{0.3cm}

$$ R_{r.m.s}^{n}= \Biggl\{ \frac{1}{Z} \lbrack
 2{<r_{1s_{1/2}}^2>}_n+ 5\frac{\sqrt{5}}{24}
 \sum_{J_cT_c(E_c)}^{}S_{J_cT_c(E_c),1p_{3/2},n}^{JT(0)}
 {<r_{1p_{3/2}}^2>}_{n,E_c} + $$

$$\hspace*{6.5cm}
 5\frac{\sqrt{10}}{24} \sum_{J_cT_c(E_c)}^{}
 S_{J_cT_c(E_c),1p_{1/2},n}^{JT(0)} {<r_{1p_{1/2}}^2>}_{n,E_c} \rbrack
 \Biggr\} ^{1/2}\ , \qquad (12b) $$
\vspace*{0.3cm}

 and for $R_{r.m.s}^m$ the expression (6с) remains valid taking into account
 (12а) and (12b).

 If the spin-orbit interaction is not taken into account, then we have:
\vspace*{0.3cm}

$$ R_{r.m.s}^{p}= \Biggl\{ \frac{1}{Z} \lbrack
 2{<r_{1s}^2>}_p\ +\ 5\frac{\sqrt{5}}{24}
\sum_{J_cT_c(E_c)}^{}(S_{J_cT_c(E_c),1p_{3/2},p}^{JT(0)}\ + $$ \\
$$\hspace*{5.0cm} \sqrt{2} \sum_{J_cT_c(E_c)}^{}
 S_{J_cT_c(E_c),1p_{1/2},p}^{JT(0)}) {<r_{1p}^2>}_{p,E_c} \rbrack
\Biggr \}^{1/2}\ ,  \ \ \ \qquad (13a) $$
\vspace*{0.3cm}

$$ R_{r.m.s}^{n}= \Biggl\{ \frac{1}{N} \lbrack
 2{<r_{1s}^2>}_p\ +\ 5\frac{\sqrt{5}}{24}
\sum_{J_cT_c(E_c)}^{}(S_{J_cT_c(E_c),1_{p3/2},n}^{JT(0)}\ + $$ \\
$$\hspace*{5.0cm} \sqrt{2} \sum_{J_cT_c(E_c)}^{}
 S_{J_cT_c(E_c),1p_{1/2},n}^{JT(0)}) {<r_{1p}^2>}_{n,E_c} \rbrack
 \Biggr\}^{1/2}\ .  \ \ \qquad (13b) $$
\vspace*{0.3cm}

 Unlike (7а) and (7в), in these expressions the dependence of the nuclear rms
 radii on the nuclear structure (the spectroscopic factors) remains.

 As the mean field, determining the one-particle levels and wave functions, in
 our work we take the potential in the form:

\setcounter{equation}{13}

\be
V(r)=V_{WS}+V_{Centrif}+V_{Coul}+V_{ls},\\
\ee

 where

$$
 V_{WS}=-V_0{ \{ 1+exp(\frac{r-r_0A^{1/3}}{a}) \} }^{-1}
$$

 -- the Woods-Saxon potential,
\vspace*{0.3cm}

$V_{Centrif}$ - the centrifugal potential,
\vspace*{0.3cm}

$V_{Coul}$ - the Coulomb potential,
\vspace*{0.3cm}

$V_{ls}=-\kappa ({\vec \sigma}\cdot{\vec l})\frac{1}{r}\frac{dV_{WS}}{dr}$  -
 the spin-orbital interaction,
\vspace*{0.3cm}

\begin{displaymath}
\vec \sigma\cdot\vec l= \Biggl \{
\begin{array}{ll}
l          & ,\ j=l+\frac{1}{2} \\
-(l+1)     & ,\ j=l-\frac{1}{2}\ ,\\
\end{array}
\end{displaymath}
\vspace*{0.3cm}

$\kappa$  - the constant of the spin-orbital interaction.
\vspace*{0.3cm}

 It is more convenient to rewrite the expressions (1) and (7) for the quadrupole
 moments as

\ba
eQ=e_p^{eff}q_p+e_n^{eff}q_n .
\ea
\vspace*{0.3cm}

{\large \bf 3. Results and discussion}\\

 The analysis of the experimental data was started from reproducing the results
 of ref. [13] and adding information on the magnetic moments.

 The proton single-particle wave functions in the nucleus $^{8}$B and
 the neutron ones in $^{8}$Li were calculated, taking into account the formula
 (10). The nucleon radius parameter $r_0$ and the surface diffuseness $a$ were
 taken as standard values: $r_0=1.27$ fm, $а=0.65$ fm. The results of
 calculations of the single-particle rms radii in accordance with expression
 (11) and with $\kappa=0$ (i.e. without spin-orbit interaction, see (14)), are
 listed in Table 1. The harmonic oscillator parameter $b=1.6$ fm was taken for
 the neutron radial wave functions in $^8$B  and the proton ones in $^8$Li
 (both for 1s- and 1p- state), like in [13].

 The results of calculations of the quadrupole and magnetic moments
 (the expressions (8), (9) and (15)) are presented in Table 2 (variant I).
 In this table the empirical values of the quadrupole moment
 $Q(^8B)$ = 6.83 fm$^2$ and $Q(^8Li)$ = 3.27 fm$^2$ [2] correspond to
 the adduced values $q_\alpha$ and $e_\alpha^{eff}$. And for the experimental
 values of the magnetic moments we have ${\mu}_{Nucl.}(^8B)$ = 1.0355 and
 ${\mu}_{Nucl.}(^8Li)$ = 1.65335 [14]. Table 2 shows that the theoretical
 ${\mu}_{Nucl.}$ are somewhat smaller than the empirical values, especially for
 $^8$Li. The values of the nuclear rms radii (the expressions (13a), (13b) and
 (6c)) are listed in Table 3 (variant I). As follows from Tables 2 and 3, we
 have reproduced the results of the calculations from [13].

 And now let us use the multiparticle shell model, where the matrix elements
 are calculated by the method of fractional parentage coefficients via
 the weight coefficients $\beta_i$ (expressions (1)$-$(5)). From
 the single-particle rms radii we will use only the values at $E_c$ = 0
 (the first line in Table 1). In the framework of this model the coefficients
 of the configurations $\beta_i$ were best fitted to the experimental values of
 $Q$ and $\mu_{Nucl.}$ for $^8$B and $^8$Li. The values of $\beta_i$ determined
 in this way are presented in Table 4 (variant A). The values $e_\alpha^{eff}$
 corresponding to these $\beta_i$ are very close to ones from [13], and good
 description is obtained for the experimental values ${\mu}_{Nucl.}$ by this
 approach (Table 2 (variant II)). The values of the nuclear rms radii, which in
 given variant of the calculations are independent from the nuclear structure,
 i.e. from the spectroscopic amplitudes (see expressions (7a) and (7b)) are
 presented in Table 3 (variant II). The comparison with the variant I in this
 table shows the considerable increase of $R_{r.m.s}^p$ for $^8$B and
 $R_{r.m.s}^n$ for $^8$Li.

 The results of calculations in the framework of two approaches require
 a detailed analysis of the considered models. We should point out
 {\it the defect} of the Cohen-Kurath and Millener-Kurath model [8] especially
 for nuclei $^{8}$Li and $^{8}$B with $J=2$ : the main component with
 $j\ =\ j^{'}\ =\ \frac{3}{2}$ in the first term of the expansion on the states
 of the residual nuclei $^7$Li and $^7$Be (expression (8))
 {\it is suppressed by chance} in this case because for $^7$Li (gr. st.) and
 $^7$Be (gr. st.) $J_c\ =\ \frac{3}{2}$, resulting to zero value of 6-j symbol,
\begin{displaymath}
\Biggl \{ \begin{array}{lll}
J    & j & J_c \\
j^{'}     & J           & 2   \\
\end {array} \Biggr \}\ =\
\Biggl \{ \begin{array}{lll}
  2    &         3/2  &        3/2  \\
 3/2   &          2   &         2  \\
\end {array} \Biggr \}\ =\ 0.
\end{displaymath}
 This component brings the dominant contribution in the calculations of
 the magnetic moments and the nuclear rms radii (expressions (9), (13а), (13в)). эта компонента
 It should be noted also somewhat {\it artificial} character of the expansion
 method on the spectroscopic factors, since the last depend both on
 the structure of the parental and residual nuclei. Moreover, using
 the expression (10) for the separation energy of the valence nucleon gives rise
 to {\it the decreasing } of the nuclear rms radii as compared with the results
 of the multiparticle shell model ( Table 3, variants I and II). This is
 a consequence of the single-particle rms radii dependence on the excitation
 energies $E_c$. If we try to reduce the expressions (13а) and (13b) to the form
 like (7а) and (7b), then it is necessary to take out of summation an average
 value of the single-particle rms radius
 ${\langle <r_{1p}^2>_\alpha \rangle}_{av.}$ indepedent on $E_c$.
 Then we obtain:

$$\hspace*{3.0cm} R_{r.m.s}^{p}={ \Biggl \{ \frac{ 2{<r_{1s}^2>}_p+
(Z-2){\langle {<r_{1p}^2>}_{p} \rangle}_{av.}} {Z}
 \Biggr \}}^{\frac{1}{2}},  \qquad\qquad\qquad (16a) $$

$$\hspace*{3.0cm} R_{r.m.s}^{n}={ \Biggl \{ \frac{2{<r_{1s}^2>}_n+
(N-2){\langle {<r_{1p}^2>}_{n} \rangle}_{av.}} {N}
 \Biggr \}}^{\frac{1}{2}},  \qquad\qquad\qquad (16b) $$
\vspace*{0.2cm}

 The average values of
 ${\langle {<r_{1p}^2>}_{p} \rangle}^{1/2}_{av.}$ = 3.575 fm and
 ${\langle {<r_{1p}^2>}_{n} \rangle}^{1/2}_{av.}$ = 3.136 fm correspond to
 the values $ R_{r.m.s}^{p}$ = 3.034 fm for $^8$B and
 $ R_{r.m.s}^{n}$ = 2.727 fm for $^8$Li, respectively. We choose the values
 $r_0$ and $a$ (the depth of the Woods-Saxon potential corresponds to
 the separation energy of the valence proton in $^8$B) so as to obtain
 ${\langle {<r_{1p}^2>}_{p} \rangle}^{1/2}_{av.}$ = 3.575 fm.
 For $r_0$ = 1.17 fm and $a$ = 0.37 fm we have obtained:
 ${\langle {<r_{1p}^2>}_{p} \rangle}^{1/2}_{av.}$ = 3.576 fm for $^8$B and
 ${\langle {<r_{1p}^2>}_{n} \rangle}^{1/2}_{av.}$ = 2.872 fm for $^8$Li.
 Given these data in the framework of the multiparticle shell model, we fitted
 the coefficients ${\beta}_i$ describing the experimental values $\mu_{Nucl.}$
 and $Q$. The values of the coefficients ${\beta}_i$ are represented in Table 4
 (variant B). The values $\mu_{Nucl.}$ and $e_{\alpha}^{eff}$ listed in Table 2
 (variant III) correspond to these ${\beta}_i$. Tables 2 and 3 (variants I
 and III) demonstrate reproduction the results of the work [13], and besides
 sufficiently good description of the magnetic moments in the given case.
 But the values $r_0$ and $a$ become much less than the standard ones and
 the nuclear wave function becomes more clustered:
 $|\beta_1|$ = 0.906  $>$ (table 4, variant B) $|\beta_1|$ = 0.859 (variant A).
 All aforesaid results were obtained with the radial one-particle wave functions
 calculated in the mean field potential without the spin--orbit interaction.
 But for such loosely-bound system as nucleus $^8B$, having the binding energy
 of the valence proton $E_{1l_j}^p\ =\ E_{1p_{3/2}}^p$ = $-$0.137 MeV,
 the probability to discover proton in the $1р_{1/2}$ state is not zero.
 And it is possible when taking into account the spin--orbit interaction
 specifically for nucleus $^8$B. In general case for a weakly-bound halo nuclei
 the value of spin-orbital interaction is an open problem at present. And in
 most cases for such nuclei, the levels of energies and the single-particle wave
 functions of $1р_{1/2}$ - states are not defined by the conventional methods,
 because the Levinson's theorem is not executed. In present work the energy and
 the wave function of the single-particle $1р_{1/2}$ - state of proton in $^8$B
 were calculated by the method of the expansion of the Sturm-Liouville
 functions, thereby taking into account the influence of the continuum [11,12].
 The constant of the spin--orbit interaction $\kappa$ was chosen so that
 the calculated single-particle energy proved to be lower than
 the centrifugal + Coulomb + spin-orbital barrier. Then we can consider this
 state with the positive energy as quasibound one. The wave function of such
 a state is convergent and can be normalized to 1. The one-particle energies
 and wave functions of the valence nucleons in $^8$B and $^8$Li were calculated
 for the ground and first excited states of the residual nuclei $^7$Li and
 $^7$Be, respectively (see expressions (10)). The spin-orbital constant
 $\kappa$ = 0.101 was chosen. Having the wave functions obtained in
 the framework of C-K and M-K model, were calculated the single-particle rms
 radii. The results of the calculations are listed in Table 5. Taking these rms
 radii (and the others represented in Table 1) we calculated the quadrupole and
 magnetic moments (the Table 2 (variant IV)) and the nuclear rms radii
 (Table 3 (variant IV)). As expected ,we obtained just a small increase of
 $R_{r.m.s}^{p}(^8B)$ and $R_{r.m.s}^{n}(^8Li)$ and the related decrease
 of $e_n^{eff}$ and $e_p^{eff}$ (as compared with the calculations without
 the spin-orbital interaction (variant I in tables 2 and 3). The results of
 calculations of the wave functions, including the spin-orbit coupling
 (see Table 5, the values for $E_c$ = 0 MeV only), in the framework of
 the multiparticle shell model are represented in Tables 2 and 3 (variants V)
 and Table 4 (variant C). The occupation numbers of the nucleon one-particle
 states are listed in Table 6. The distributions of the proton density in $^8$B
 (solid curve) and the neutron density in $^8$Li (dashed curve) are shown
 in Fig.1. In this figure the dash-dot curve corresponds to the distribution of
 the neutron density in $^8$B or the proton density in $^8$Li, as far as in both
 cases we used the harmonic oscillator wave functions with the equal values of
 the oscillator parameter. The comparison with the results of the calculations
 of the wave functions in the mean field potential without the spin--orbit
 interaction (variants II in tables 1 and 2, variant A in table 4) shows only
 a slight increase in $R_{r.m.s}^{p}(^8B)$ and $R_{r.m.s}^{n}(^8Li)$. It is
 caused by the small value of the spin-orbit potential $V_{ls}$.
 $V_{ls}$ = 4 MeV corresponds to the value $\kappa$ = 0.101, while $V_{ls}$
 varies within 8 $\div$ 13 MeV for the stable nuclei of $1р$-shell [15].
 The values of the single-particle rms radii in $^8$B
 ${<r_{1p_{3/2}}^2>}_{p}^{1/2}$ = 4.360 fm and
 ${<r_{1p_{1/2}}^2>}_{p}^{1/2}$ = 4.677 fm (Table 5) are significantly
 smaller than the corresponding value ${<r_{1p_{3/2}}^2>}_{p}^{1/2}$ = 6.83 fm,
 calculated by the quasiparticle random phase approximation model [16],
 but larger than ${<r_{1p_{3/2}}^2>}_{p}^{1/2}$ = 4.02 fm $-$ a pure mean-field
 plus pairing correlation calculation [16]. It follows from Table 3
 (variants II and V) and Fig. 1 that nucleus $^8$B has a significant proton
 halo, and $^8$Li has a less neutron halo. And in this case, a neutron (proton)
 in $^8$B ($^8$Li) is found in $1р_{1/2}$ - state with a low probability
 $\sim$12$\%$, whereas one of the three 1p-protons(neutrons) in $^8$B ($^8$Li)
 can be found in this state already with a higher probability $\sim$46$\%$
 (Table 6). Our values of the nuclear rms radii, calculated in the framework of
 the multiparticle shell model exceed those available in the literature
 including [7] (Table 3, variant VI). The authors of [7] reproduced the binding
 energy of the valence proton and the large quadrupole moment of $^8$B, using
 the microscopic multicluster model, however they could not describe
 the magnetic dipole moments of $^8$Li and $^8$B (Table 2, variant VI).
 Of course, in spite of the successful description of the magnetic and
 quadrupole moments of the mirror nuclei $^8$Li and $^8$B in the framework of
 our approach, the obtained results are preliminary yet. And it would be very
 desirably to have the experimental data of the electron scattering on these
 nuclei and also the data of the polarization measurements. As far as
 the electron scattering on unstable nuclei has not been realized yet, it would
 be actual to carry out experiments on scattering of secondary beams of $^8$Li
 and $^8$B on polarized protons.\\

{\large \bf 4. Conclusion}\\

 We used the multiparticle shell model with:
 the radial wave functions, calculated in a realistic Woods-Saxon potential;
 the exact taking into account of a continuum;
 the use of the method of fractional parentage coefficients --
 for analysis of the experimental electric quadrupole and magnetic dipole
 moments of nuclei $^8$Li and $^8$B. The obtained values of the nuclear rms
 radii and the nucleon density distributions show that nucleus $^8$B has
 a significant proton halo, and $^8$Li has a less neutron halo. We compared
 applicability of the multiparticle shell model and the microscopic multicluster
 model to the weakly bound halo nuclei. For the obtaining more correct wave
 functions of the considered nuclei, it is necessary to have the experimental
 scattering data of secondary beams of $^8$Li and $^8$B on polarized protons.

\newpage
}
\begin{center}
\hspace*{-2.cm}{ Bibliography} \\
\end{center}

1. B. Blank, C. Marchland, M.S. Pravikoff et al.//
 Nucl. Phys. 1997. V. A 624. P. 242.\\

2. T. Minamiosono et al. // Phys. Rev. Lett. 1992. V. 69, No 14. P. 2058.\\

3. I. Tanihata. // Nucl. Phys. 1995. V. A588. P. 253.\\

4. J.A. Wheeler. // Phys.Rev. 1937. V. 52. P. 1083, P. 1107;\\
\hspace*{1.0cm} D.A. Hill, J.A. Wheeler. // Phys. Rev. 1953. V. 89. P. 1102.\\

5. K. Vildermut, Y. Tan. // "The General Theory of The Nucleus".  "Mir", Moscow 1980. \\

6. V.G. Neudachin, Y.F. Smirnov. // \\
\hspace*{1.0cm} "Nuclear associations in light nucleus".  "Nauka", Moscow 1969.\\
\hspace*{1.0cm} O.F. Nemec, V.G. Neudachin, A.T. Rudchik, Y.F. Smirnov, Y.M. Chuvil'skii. // \\
\hspace*{1.0cm} "Nuclear associations in the atomic  nucleus  and
nuclear reactions \\ \hspace*{1.0cm} of the
 nucleus-nucleus collisions". "Naukova Dumka", Kiev 1988.\\

7. K. Varga, Y. Suzuki and I. Tanihata.//Phys.Rev. 1995. V. C52. P. 3013; \\
\hspace*{1.0cm} K. Varga, Y. Suzuki, K. Arai and Y. Ogawa. //
Nucl. Phys. 1997. V. A616. P. 383.\\

8. S. Cohen and D. Kurath. // Nucl. Phys. 1965. V. 73. P. 1; \\
\hspace*{1.0cm} D.J. Millener and D. Kurath. // Nucl. Phys. 1975. V. A225. P. 315.\\

9. S.M. Bekbaev, G. Kim, R.A. Aramjan. // \\
\hspace*{1.0cm} Izv. AN USSR. Сер. физ. 1990. Т. 54. С. 1014.\\

10. F.A. Gareev, G.  Kim,  A.V.  Khugaev.  // \\
\hspace*{1.0cm} Yad Phiz.  [Russian J.Nucl.Phys] 1997. No. 12. P. 1240.\\

11. E. Bang, F.A. Gareev, S.P. Ivanova. // ACHAYA. 1978. V. 9, N. 2. P. 286.\\

12. Il-Tong Cheon, G. Kim, A.V. Khugaev.//
 Prog. Theor. Phys. 1998. V. 100, N. 2. P. 327. \\

13. H. Kitagawa and H. Sagawa. // Phys. Lett. 1993. V. B299. P. 1. \\

14. F. Ajzenberg-Selove. // Nucl. Phys. 1984. V. A413. P. 1. \\

15. L.R.B. Elton and A. Swift. // Nucl. Phys. 1967. V. A94. P. 52. \\

16. W. Schwab, H. Geissel, H. Lenske et al. // Z. Phys. 1995. V. A350. P. 283.  \\

\newpage

{\bf TABLES \\ }

\vspace*{1.5cm}
{\bf Table 1.}
 The one-particle root-mean-square (rms) radii for the
 proton one-particle state in    $^8B$ and  the
 neutron one-particle state in    $^8Li$   \\

\vspace{2mm}
\begin{center}
\begin{tabular}{|c|c||c|c|}  \hline \hline
$E_c({}^7Be)$, [MeV]    & ${<r_{1p}^2>}_{p,E_c}^{1/2}(^8B)$, [fm] &$E_c(^7Li)$, [MeV] & ${<r_{1p}^2>}_{n,E_c}^{1/2}(^8Li)$, [fm] \\ \hline \hline
 0.0    & 4.359  & 0.0      & 3.620   \\
 0.4292 & 3.911  & 0.4776   & 3.463   \\
 4.5700 & 2.923  & 4.6330   & 2.851   \\
 6.7300 & 2.750  & 6.6800   & 2.709   \\
 7.2100 & 2.719  & 7.4670   & 2.666   \\
 9.2700 & 2.610  & 9.6100   & 2.560   \\
 9.9000 & 2.582  & 10.250   & 2.541   \\
 11.010 & 2.538  & 11.250   & 2.505    \\ \hline \hline
\end{tabular} \\
\end{center}

\newpage

\hspace*{1.5cm} {\bf Table 2.} Matrix elements $q_\alpha$ (fm$^2$),

\hspace*{1.5cm} the empirical effective charge of nucleon
$e_{\alpha}^{eff}$ and theoretical
magnetic moments,

\hspace*{1.5cm} (experimental   $\mu_{Nucl.}(^8B)$ = 1.0355 and
$\mu_{Nucl.}(^8Li)$ = 1.65335 [14] ).

\vspace{2mm}
\begin{center}
\begin{tabular}{|c|c|c|c|c|c|c|}  \hline \hline
Variant  & Nuclei &$q_n$& $q_p $&  $e_{n}^{eff}$ &  $e_{p}^{eff}$ &   $\mu_{Nucl.}$  \\
$$       &      &$   $& $    $&  $           $ &  $           $ &   $ $ (theor.)     \\ \hline  \hline
       & ${}^8Li$  & 4.163  & 0.800   &       &        & 1.367   \\
    I  &           &        &         & 0.577 & 1.095  &         \\
       & ${}^8B$   & 0.800  & 5.813   &       &        & 0.988   \\ \hline
       & ${}^8Li$  & 3.914  & 0.936   &       &        & 1.604   \\
    II &           &        &         & 0.570 & 1.109  &         \\
       & ${}^8B$   & 0.936  & 5.677   &       &        & 1.021   \\  \hline
       & ${}^8Li$  & 3.592  & 1.107   &       &        & 1.584   \\
   III &           &        &         & 0.567 & 1.114  &         \\
       & ${}^8B$   & 1.107  & 5.567   &       &        & 1.016   \\   \hline
       & ${}^8Li$  & 4.574  & 0.800   &       &        & 1.367   \\
    IV &           &        &         & 0.537 & 1.022  &         \\
       & ${}^8B$   & 0.800  & 6.263   &       &        & 0.988   \\    \hline
       & ${}^8Li$  & 4.024  & 0.879   &       &        & 1.603   \\
     V &           &        &         & 0.570 & 1.109  &         \\
       & ${}^8B$   & 0.879  & 5.705   &       &        & 1.020   \\ \hline
       & ${}^8Li$  & 4.024  & 0.879   &       &        & 1.603   \\
     V &           &        &         & 0.570 & 1.109  &         \\
       & ${}^8B$   & 0.879  & 5.705   &       &        & 1.020    \\  \hline
       & ${}^8Li$  &  $-$   &  2.23   &       &        & 1.42    \\
 VI [7]&           &        &         &  $-$  &  $-$   &         \\
       & ${}^8B$   &  $-$   &  6.65   &       &        &  1.17   \\ \hline \hline
\end{tabular} \\
\vspace*{0.3cm}

{\footnotesize Attention. In the work [7]
effective charges have not used.

 т.е. $e^{eff}_n$ = $e_n$ = 0, $\ e^{eff}_p$ = $e_p$ = 1.}
\end{center}
\newpage

\begin{center}
{\bf Table 3.} The proton, neutron and matter rms radii. (fm)

\vspace{5mm}
\begin{tabular}{|c|c|c|c|c|}  \hline \hline
Variant & Nucleus & $R_{r.m.s}^{p}$ &$R_{r.m.s}^{n}$ & $R_{r.m.s}^{m} $    \\ \hline
& ${}^8B$ & 3.034 & 2.164 & 2.740 \\
I  &   &   & &\\
& ${}^8Li$ & 2.164 & 2.727 & 2.531 \\ \hline
           & ${}^8B$ & 3.597 & 2.164 & 3.138 \\
 II &   &   &   &\\
  & ${}^8Li$ & 2.164 & 3.065 & 2.763 \\ \hline
   & ${}^8B$ & 3.034 & 2.164 & 2.741 \\
 III  &   &   & &\\
        & ${}^8Li$ & 2.164 & 2.547 & 2.411 \\ \hline
   & ${}^8B$ & 3.082 & 2.164 & 2.775 \\
 IV  &   &   & &\\
        & ${}^8Li$ & 2.164 & 2.854 & 2.617 \\ \hline
   & ${}^8B$ & 3.634 & 2.164 & 3.164 \\
 V  &   &   & &\\
        & ${}^8Li$ & 2.164 & 3.079 & 2.772 \\ \hline
   & ${}^8B$ & 2.830 & 2.260 & 2.630 \\
 VI [7]  &   &   & &\\
        & ${}^8Li$ & 2.190 & 2.600 & 2.450 \\ \hline
\end{tabular} \\
\end{center}

\vspace*{1.0cm}

\begin{center}
{\bf Table 4.}  The weights of the corresponding configurations

\vspace{5mm}
\begin{tabular}{|c|c|c|c|}  \hline \hline
Configuration  & \multicolumn{3}{|c|}{$\beta_i$} \\ \cline{2-4}
&  Variant A &   Variant B & Variant C \\ \hline
$[31]{}^{33}P$& -0.859 & -0.906 &-0.850  \\
$[31]{}^{31}D$&  -0.378  &  -0.269  &    -0.388    \\
$[31]{}^{33}D$& -0.241  &  -0.185  &  -0.241  \\  $[31]{}^{33}F$&
-0.124 & -0.069 & -0.131 \\ $[22]{}^{33}D$&  -0.193  &  -0.236  &
-0.202  \\  $[211]{}^{33}P$&  -0.055  &  0.049  &    -0.065    \\
$[211]{}^{35}P$&  -0.072  &  0.093  &  -0.085  \\  \hline  \hline
\end{tabular} \\
\end{center}

\newpage

\begin{center}
{\bf Table 5.} Energies and root-mean-square states
(energies (MeV), radii (fm) ).

\vspace{5mm}
\begin{tabular}{|c|c|c|c||c|c|c|c|}  \hline \hline
\multicolumn{4}{|c|}{$^8B$} & \multicolumn{4}{|c|}{$^8Li$}  \\ \hline
$E_c(^7Be)$  & $1l_j$ & $E_{1l_j}^p$ & $<r_{1l_j}^2>_p^{1/2}$ &
$E_c(^7Li)$  & $1l_j$ & $E_{1l_j}^n$ & $<r_{1l_j}^2>_n^{1/2}$ \\ \hline
& $1p_{3/2}$ & -0.137 & 4.360 & &$1p_{3/2}$ & -2.000 & 3.521 \\
0.0 & & & & 0.0 & & &  \\
    & $1p_{1/2}$ & 1.004 & 4.667 & &$1p_{1/2}$ & -0.832 & 4.229 \\ \hline
    & $1p_{3/2}$ & -0.566 & 3.888 & &$1p_{3/2}$ & -2.4776& 3.380 \\
0.4292 & & & & 0.4776 & & &  \\
& $1p_{1/2}$ &  0.648 & 4.172 & &$1p_{1/2}$ & -1.234 & 3.875 \\ \hline \hline
\end{tabular} \\
\end{center}

\vspace*{2.0cm}


\begin{center}
{\bf Table 6.} The occupation numbers of the nucleon $1p_j$ $-$
states in $^8B$.

\vspace{5mm}
\begin{tabular}{|c|c||c|c|}  \hline \hline
 $ $                                     & $ $                                      & $ $                                      & $ $                                        \\
$\qquad {\eta}_{1p_{3/2}}^{(p)} \qquad $ & $\qquad {\eta}_{1p_{1/2}}^{(p)} \qquad $ & $\qquad {\eta}_{1p_{3/2}}^{(n)} \qquad $ & $\qquad {\eta}_{1p_{1/2}}^{(n)} \qquad $   \\
 $ $                                     & $ $                                      & $ $                                      & $ $                                        \\  \hline  \hline
 $ $  & $ $    & $ $   & $ $   \\
 2.543& 0.457  & 0.878 & 0.122 \\
 $ $  & $ $    & $ $   & $ $   \\ \hline \hline
\end{tabular} \\
\end{center}

\newpage


\parbox[h]{16.cm}{\it Fig. 1.
The distributions of the proton density in $^8$B
 (solid curve), \\
 the neutron density in $^8$Li (dashed curve) \\
the neutron density in $^8$B or the proton density in $^8$Li
(the dash-dot curve). \\ }


\end{document}